\documentstyle[astron,12pt]{l-aa}
\newcommand{\EQ}{\begin{equation}}
\newcommand{\EN}{\end{equation}}
\newcommand{\EQA}{\begin{eqnarray}}
\newcommand{\ENA}{\end{eqnarray}}

\input psfig

\begin{document}
\thesaurus{11.05.2, 11.17.3, 12.03.2}
\title{ $V/V_{max}$ Statistics and Neo-Classic Cosmological Tests }


\author{ L. Van Waerbeke$^1$, G. Mathez$^1$, Y. Mellier$^1$, H. 
Bonnet$^1$, M. Lachi\`eze--Rey$^2$}

\offprints{L. Van Waerbeke}
\institute{  $^1$ Laboratoire d'Astrophysique de Toulouse, URA285.
Observatoire Midi-Pyr\'en\'ees 
14, avenue Edouard Belin F-31400 - Toulouse
$\;\; ^2$ CEN Saclay Service de Physique Th\'eorique F-91191 Gif Cedex
}
\date{Received ; Accepted }
\maketitle

\begin{abstract}
A new cosmological test is derived, based on 
the distribution of individual $V/V_{max}$ in 
a complete redshift--limited sample of distant objects.
The fundamental assumption is that, in any range of absolute 
luminosity, individual 
$V/V_{max}$ are required to be uniformly spread over the $[0,1]$ range.
Under the assumption of Pure Luminosity Evolution,
this gives rise to a natural partition of the sample into
{\it high luminosity, redshift-limited} and {\it low luminosity 
magnitude--limited} quasars. The behavior of $V/V_{max}$ versus evolution 
and cosmology differs substantially in the two subsamples. 
This condition of uniformity is 
probed in any absolute magnitude bin, allowing a likelihood function
to be computed from the Kolmogorov--Smirnov probabilities of each bin. 
Monte--Carlo simulations show that the test is mostly sensitive 
to the density parameter, but, under certain conditions, it also sets 
constraints on the space curvature and, to a lower extent, on the 
cosmological constant. The efficiency of this test applied to 
two kinds of simulated quasar samples 
is examined: large number QSO sample, but limited to 
redshifts $z<2.2$ or smaller in QSO number, but with higher a redshift limit.
Power law and exponential luminosity evolution
laws are compared, and cross--tests show that the functional form 
of luminosity evolution does {\it not} affect substantially the
probabilities in the parameter space $(\Omega_{mat}, \Lambda)$.
The test is then applied to the UVX sample of Boyle et al. (1990). 
A low matter density, and a flat Universe without cosmological constant, are
rejected:
$0.2<\Omega <0.8 $ within the 95 \% confidence level. 
\end{abstract}
\section{Introduction}

At the end of the sixties, the discovery of quasars was a promising tool
for cosmology.
In particular $<V/V_{max}>$, originally designed to test the spatial
uniformity of a stellar population, was used as a cosmological test.
This test was expected to reject all of the 'bad' cosmologies where the average
$<V/V_{max}>$ differs significantly from 1/2. 
The first application to quasar samples lead to 
$<V/V_{max}>$ significantly above 1/2 (Schmidt 1968), {\it whatever 
the assumed cosmology}. This finding was interpreted either in terms of 
a strong {\it Density Evolution (DE)} (Schmidt 1972), or in terms of 
{\it Luminosity evolution} (Mathez 1976, 1978). 
The $<V/V_{max}>$ test is still the clearest evidence for quasar
evolution (however, see Bigot and Triay 1991). 

The situation soon 
appeared to be quite hopeless for 
the cosmological tests with QSOs: whatever the cosmology, 
whatever the nature and the functional form assumed for evolution,
there exists a value of the characteristic evolution parameter ensuring
$<V/V_{max}>=1/2$ within the statistical errors. In spite of many attempts, no 
reliable theoretical evolutionary model has been derived so far from the
physics of quasar emission. Consequently, the study of evolution
remains essentially phenomenological. Several papers 
followed similar procedures (Schmidt 1972; Mathez 1976;
Marshall {\it et al.} 1983) in order to associate a value of the 
evolution parameter to a given cosmology. 
Turner (1979) proposed a
slightly different way to perform magnitude-redshift tests 
with quasars.
A Luminosity Dependent DE model (LDDE) was also 
proposed (Green et Schmidt 1983); 
this model however requires more data
because it depends on two parameters at least.
Applying $<V_e/V_{a}>$ tests (in the sense of Avni and Bahcall 1980) 
to a composite sample of 300 objects, 
Kassiola and Mathez (1991) found that no single parameter 
evolution law (neither DE nor LE) gives complete satisfaction. 
This was probably due to sample inhomogeneity, since 
(Boyle {\it et al.} 1987,1990) found that
a single, large, complete sample of 400 UVX QSOs 
is consistent with {\it Pure Luminosity Evolution (PLE)}.

Several new cosmological tests using QSOs have been recently proposed.
Schade and Hartwick (1994) assume a functional form for LE
and perform the Loh-Spillar test with QSOs (Loh and Spillar 1986). 
Deng {\it et al.} (1994)
find a typical scale of the order of 100 Mpc in the spatial distribution
of QSOs, and derive a cosmological test by comparing it to preferred
scales in the galaxy
spatial distribution (Broadhurst {\it et al.} 1990). However there is 
no clear consensus on this subject, since Shanks and Boyle (1994) also
claim to detect significant correlation, but on scales one order 
of magnitude smaller. Phillipps (1994) 
applies a test originally proposed by Alcock and Paczynski (1979) to measure
the cosmological constant from the mean projection angle between
quasar-quasar separation and the line of sight. Finally Malhotra and
Turner (1995) study the evolution of quasars in cosmological constant
dominated flat universes.

This paper is the first of a series related to a new
cosmological test using a neo--classical $<V/V_{max}>$ approach. 
The motivation was to address the problem of evolution of QSOs in non-zero 
$\Lambda$ cosmologies, where evolution should be less.
(Mathez {\it et al.} 1995, Paper II)
Previous analyses did not consider this case, except one paper (Malhotra 
and Turner 1995). However, some recent determinations of the Hubble 
constant from HRCam images (Pierce {\it et al.} 1994) or HST images (Freedman 
{\it et al.} 1994) give high values of $H_0$ incompatible with the age of globular 
clusters, unless $\Lambda >0$ (see e.g. the discussion in 
Leonard and Lake 1995). On the other hand, the statistics of 
multiply imaged QSOs (Kochanek 1990; Helbig {\it et al.} 1995), as well 
as galaxy number counts from Yoshii {\it et al.} (1995) provide an upper 
limit on $\Lambda$, which seems lower than $0.9$. If it is confirmed that 
$\Lambda>0$, then the discrepancy between the vacuum energy of physicists 
and the cosmological constant of astrophysicists could be a severe problem 
for standard physics. The alternative theory of Scale Relativity 
(Nottale, 1993) which introduces $\Lambda ^{-1/2}$ as a natural limiting 
length analogous to the velocity of light in Special Relativity, should be 
explored seriously.

The present paper is exclusively concerned with PLE, 
and a forthcoming paper will be devoted to PDE. 

In Section 2, the basic ideas of the test and the fundamental
hypotheses on QSO statistics are given.
In Section 3, we show how to associate a probability to a cosmological
model. 
In Section 4, we derive the expression of $V/V_{max}$ suited for the
brightest quasars. 
Catalogues are produced with Monte--Carlo simulations in Mathez {\it et
al.} 1995, (hereafter Paper II). The test is
applied to these catalogs in Section 5, and on a real QSO catalogue in
Section 6.
The main results are discussed in the last Section, where the hypotheses are
summarized.

\section{ Distribution of Maximum Volumes in a Complete Sample}

\def\col#1#2{\pmatrix{#1\cr \vbox to 2mm{} \cr #2 \cr}}
\subsection{The spirit of the test}
Our working frame is the Friedmann-Lema\^ \i tre cosmologies.
The QSO Luminosity Evolution is described, as usual, 
by a given functional form depending on a single evolution parameter.
Our null hypothesis is that the QSOs are uniformly distributed in
volume, and that we deal with a sample complete in magnitude.
To properly account for the well known observational biases present
in complete QSO samples at low and high redshifts,
one has to work in a restricted redshift range.
The variable which is uniformly distributed is not $V/V_{max}$, but 
a ratio $x$ defined in Section 4.
The sample is binned according to $M_0$, the absolute magnitude
defined at the epoch $z=0$ and at the
frequency where observations are done. 
Since in the frame of the correct model (evolution and cosmology) the
$x$ distribution does not depend on the absolute magnitude, 
each bin must have a uniform $x$ distribution.  A correlation 
between $V/V_{max}$ and absolute magnitude is one of the most sensitive 
pieces of evidence for a wrong choice of cosmology and/or 
evolution (see Fig. 1 and Bigot and Triay 1991).
Fig. 1 shows the positions of QSOs in the $(M_0,V/V_{max})$
plane for three values of the evolution parameter. 
A study of the $V/V_{max}$ distribution
is clearly less sensitive in the whole sample than in luminosity bins.

\begin{figure}
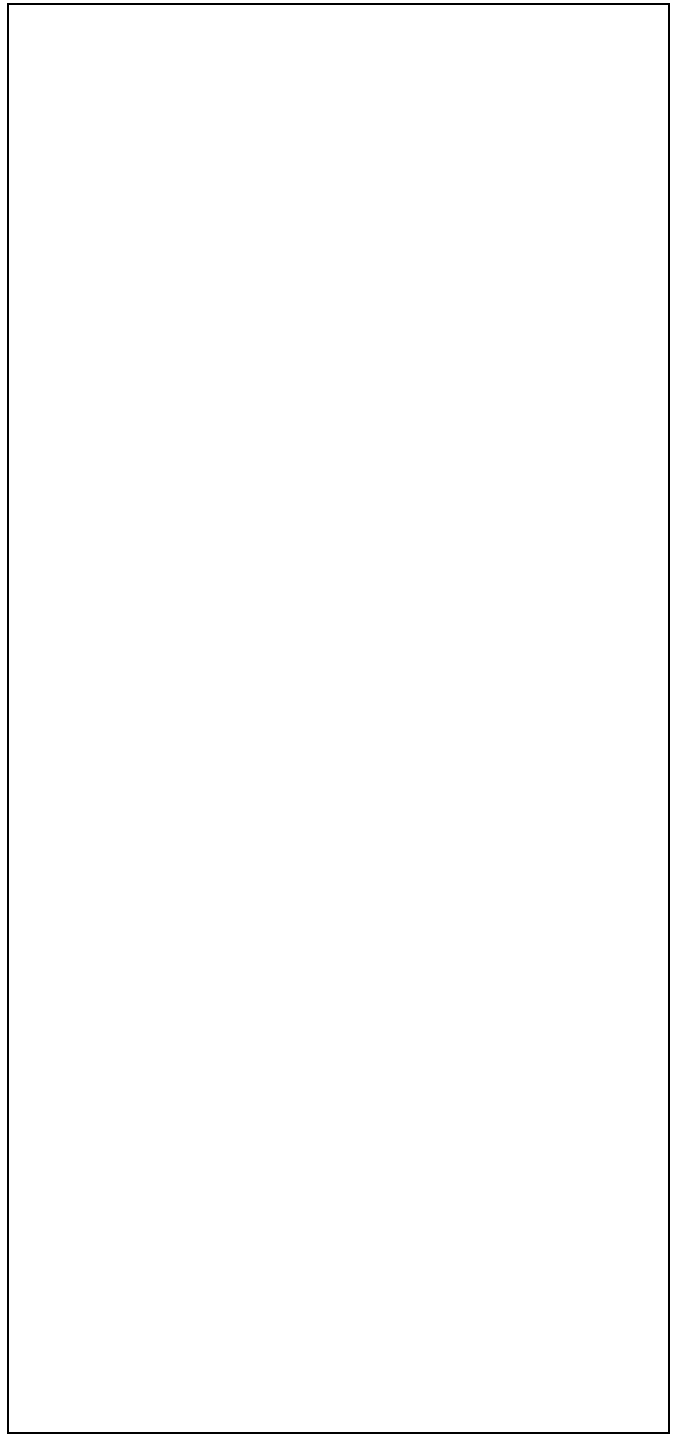

\picplace{19cm}
\caption{The basic idea of the test: Boyle's et al. (1990)
QSOs are represented in the (absolute magnitude-$V/V_{max}$) plane. 
The model is power law luminosity evolution
(PWLE), $\Omega=1, \Lambda=0$. The values of the evolution parameter are
$k_L=0., 3., 6.$, respectively. The upper and lower plots show a strong
correlation, corresponding to a bad combination of evolution and 
cosmological model.
In the middle, there is no correlation whatever the magnitude.
It can be seen that the {\it global} $<V/V_{max}>$ may be 1/2, while 
the distribution is far from uniform in the high-- and low--luminosity
bins.}
\end{figure}

Another interesting piece of information is the evolution of individual $x$ 
versus the cosmology and the evolution characteristic parameters (see Fig. 2). 
We expect that a cosmological test using the $x$ values might be mostly 
sensitive in the regions of parameter space where $x$ vary strongly. As we see 
in Fig. 2, two asymptotic regions, toward $\Lambda <0$ and $\Omega_{mat}>1$, are
observed, which show that the test will be insensitive to the cosmology in
these regions.

\begin{figure}
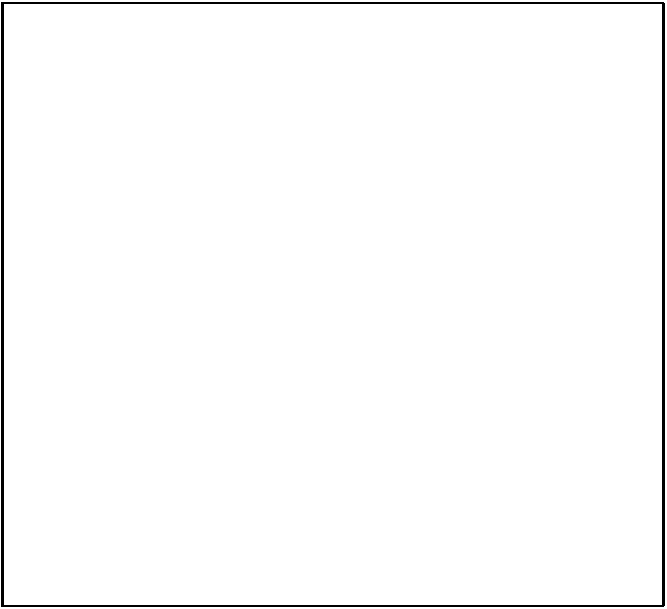

\picplace{8cm}
\caption{$V/V_{max}$ ratio versus cosmological parameters for
15 QSOs randomly chosen in the Boyle's {\it et al.} (1990) sample.
Top left, $\Lambda=0$, top right, $\Lambda=1$, bottom left,
$\Omega_{mat}=1$, bottom right, $\Omega_{mat}=2$. 
The evolution parameter always corresponds to the maximum likelihood.
Discontinuities in the derivative correspond to change of status from
magnitude-- to redshift--limited or conversely. Noise in the bottom
curves is due to computational difficulties. In the two asymptotic regimes
discussed in text (towards large $\Omega $ and small $\Lambda $),
the test has little or no sensitivity, while the test is quite efficient at
low $\Omega $.}
\end{figure}

\subsection{ Luminosity Evolution}
Pure Luminosity Evolution has been introduced in a phenomenological way,
without physical support. It corresponds to the case where the fraction
of active galaxies is constant versus redshift, but not their luminosity. 
Although it is a statistical evolution of the whole population,
the usual simplifying hypothesis on luminosity evolution is that {\it all}
quasar absolute luminosities $L$ do evolve in parallel according to the law
(Mathez 1976, Marshall et al. 1984, Boyle et al. 1987):
\EQ
L\left( t(z) \right) = L( t_0) \times e\left(t(z)\right),
\EN
where $e(z)$ has either the 'power law' form in the PWLE case:
\EQ
e(z) = (1+z)^{k_L},
\EN
or the 'exponential' form in the LEXP case:
\EQA
e(z)&=&exp \left[\left(t_0-t(z)\right)/\tau \right]\nonumber\cr
&=&exp \left[k_L H_0\left(t_0-t(z)\right)\right].
\ENA
$t(z)$ is the cosmological epoch corresponding to redshift $z, t_0 =
t(z=0)$, $\tau =1/(k_L H_0) $ is the characteristic evolution time,
and $k_L$, the 'evolution strength' in the standard notation.

\subsection{Evolutionary tracks in the Hubble plane}
Several pairs of independent variables may be used to 
study the quasar distribution: redshift, volume, and comoving volume on 
one hand and apparent magnitude, flux density, intrinsic luminosity 
on the other hand. Here we prefer to work in the Hubble plane,
apparent magnitude--redshift ($m,z$), because it corresponds to observables. No
hypothese either on cosmology or on evolution are needed to represent 
quasars in this plane. 
 
Assume that we have a quasar sample complete 
in the magnitude range $[m_1,m_2]$, in a given sky area. 
The fainter limit $m_2$ is obvious. The bright limit $m_1$
may arise from photometric saturation (either in the photometric or in
the spectroscopic data). 
Quasars of the sample thus lie inside what we call 'box $\cal B$'
in the following, i.e. the domain
$[m_1,m_2]\otimes [z_1,z_2]$ of the Hubble plane:
we further restrict ourselves to some range 
of absolute magnitude $[M_0^1,M_0^2]$, arbitrary provided that 
$[M_0^1,M_0^2]\otimes[z_1,z_2]$ is included in box $\cal B$.

According to Eq. 2, all quasars follow parallel
evolutionary tracks in the Hubble plane.
Now consider a given quasar in box $\cal B$.
Its evolutionary track will go along the 
$m(z)$ curve in the Hubble plane, described by the Mattig
relation:
\EQ
m(z)=M_0-5+2.5log_{10}\left( d_L(z)^2 \times K(z) /e(z) \right)
\EN
where $M_0$ is the absolute magnitude at epoch $z=0$, 
$d_L(z)$ is the luminosity distance and $K(z)$ is the
K--correction: $K(z)=(1+z)^{\alpha -1}$ with $\alpha =0.5$ as spectral
index. Note that in the case of a strong luminosity evolution, the $m(z)$
relation is no longer monotonic, and can even exhibit several
local maxima in some extreme cases.
It is necessary to exclude for any volume computation, all 
the redshift ranges where $m>m_2$ or $m<m_1$ (see the details in Paper II). 
The stronger the evolution, the smaller will
be the slope of the evolutionary track. It can become negative if evolution
is such that apparent magnitude passes through a maximum.
Fig. 3 shows one illustration of the Mattig relation when the evolution
is stronger than the cosmological extinction.

\begin{figure}
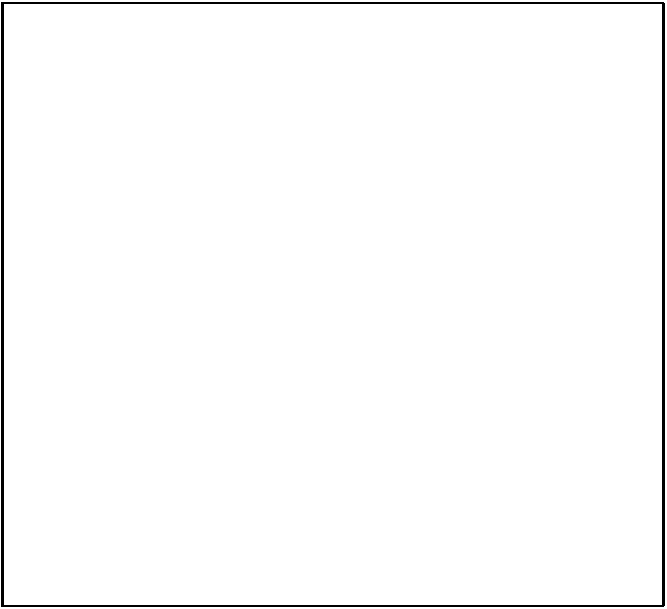

\picplace{8cm}
\caption{Various Mattig functions with different absolute magnitudes,
show parallel evolutionary tracks in the Hubble plane. 
Between the separators $\cal S_1$ and $\cal S_2$, all QSOs are 
redshift--limited.
The sign $+$ on the bottom line denotes a QSO whose
redshift $z$ is limited in the range $[z_{min},z_{max}]$ because of
the bright magnitude limitation $m<m_1$.}
\end{figure}

\subsection{Observational cutoffs at low and high redshift}
As has been known for a long time, the redshift
distribution of most large complete QSO samples is biased at low and high
redshift. These cutoffs are
important to account for, since they govern the locus of both
extremities of evolutionary tracks in the Hubble plane.

At low redshift, it is easier to distinguish the
host galaxy of QSOs, so many of them {\it do not} have a stellar 
appearance and are missed in quasar samples. 
We will restrict our analysis to redshifts z$>$0.3.

Searching for ultraviolet excess (UVX) stellar objects 
is the most efficient way to detect
quasar candidates, so the largest complete
 samples consist of UVX quasars.
At high redshift, quasars no longer have any 
ultraviolet excess, since 
$Ly_{\alpha }$, the dominant emission line, passes to the blue
filter at $z\simeq 2.3$.
So most of the large quasar samples are in fact limited to $0.3<z<2.2$. 
To our knowledge, Kassiola and
Mathez (1991) and Crawford (1995) are the first authors to explicitly 
mention that using maximum or minimum redshifts outside the range $[0.3,2.2]$ 
for UVX QSOs is erroneous.
In Paper II, the joint distribution of redshift and magnitude 
is examined in the frame of PLE, and it is shown how to compute the
individual ratios $x$ which are uniformly distributed over $[0,1]$ for
quasars in any arbitrary bin of absolute magnitude.

In the next Section, we will take advantage of these properties to derive a
cosmological test.

\section{ Maximum Volume and Minimum Entropy }

Our improved $V/V_{max}$ analysis,
essentially based on the Kolmogorov--Smirnov test, 
is somewhat similar to that used in the study by Zieba and Chyzy (1991) 
of high-- and low--redshift radio galaxies.
We show with details hereafter how it 
enables us to associate a significance level to each cosmological model
and to reject some of them.

\subsection{ Choosing a grid of cosmological models}
Most of the previous cosmological tests have been
performed under the assumption $\Lambda =0$, with either a low
density parameter $\Omega _{mat}$ or a flat universe $\Omega _{mat}=1$
($\Omega _{mat}$ is the {\it total} density
parameter, including baryonic and non-baryonic matters).
Some tests however adopt a non zero cosmological constant 
but impose a null curvature by the condition $\Omega _{mat}+\Lambda =1$. 
Present ideas on the density parameter are synthetized in Table 4.
To explore a large range of cosmological models, we choose the range
$[0,3]$ for $\Omega_{mat}$, and $[-1,2]$ for $\Lambda $, varying these
parameters independently, by steps of 0.2. However we exclude from the
test all cosmologies without an initial singularity (see Carroll {\it et
al.} (1992) for details).
It results in 16$\times $16-17=239 cosmological models.
Absolute magnitudes are irrelevant in the test, only the
rank of these magnitudes is important, which is 
independent of $H_0$. Our test, based on the statistics of individual
$V/V_{max}$ ratios, is {\it totally independent} of the Hubble
parameter. 

\subsection{Associating a confidence level to each cosmological model } 
We work in the $(absolute \; magnitude - V/V_{max})$
plane. We choose a functional form of the evolution before performing
the test.
Our test consists in ensuring a maximum likelihood to the null
hypothesis of {\it uniform distributions} of the $x$ ratio
defined in Section 4 {\it jointly in all bins of
absolute magnitude}. We have to ensure this {\it without any hypothesis on the
functional form of the luminosity function}.

Let ${\cal M}$ be the current cosmological model $(\Omega_{mat}, \Lambda)$.
To assign a probability to ${\cal M}$, we proceed as follows:
once a value of the evolution parameter $k_L$ is chosen, each ratio $x$
is computed.
The sample is ordered by increasing absolute
luminosity and divided into $n$ luminosity
bins containing equal numbers of quasars. This is to minimize
the noise that would be caused by very different numbers 
of quasars in each bin. 
In each luminosity bin we calculate the Kolmogorov-Smirnov probability
that the corresponding $x$ ratios come from a uniform
distribution (see Press et al. 1992). Let
$P_{KS}^i({\cal M},k_L)$ be this probability for the luminosity bin 
 $i$. These bins are statistically independent, so are the
probabilities, thus we can compute the likelihood function of 
the total sample of $x$:
\EQ
{\cal L} ({\cal M},k_L)=\prod_{i=1}^{N} P_{KS}^i({\cal M},k_L)
\EN
This likelihood is used to calculate an entropy:
\EQ
{\cal E}=-2log({\cal L})
\EN
For each model ${\cal M}$, we determine the parameter
$k_L$ giving the lowest value of the entropy ${\cal E}_{low} (\cal M)$.
${\cal E}_{low} (\cal M)$ is defined as the entropy of the cosmological
model $\cal M$.  This operation is repeated for the 239
cosmological models to get an entropy map. At the end, the entropy is 
${\cal E}_{low}({\cal M})={\cal E}_{min}+\Delta {\cal E}({\cal M})$, 
where ${\cal E}_{min}$ is the lowest entropy found in the 
cosmological grid. $\Delta {\cal E}$ follows a $\chi^2$
statistic with three degrees of freedom (Lampton {\it et al.} 1976), 
since three parameters are determined. For three parameters, the confidence 
intervals around the
best model $\cal M$ are given by $\Delta {\cal E}=6.25, 7.80, 11.30,
16.20$ for the $90\%, 95\%, 99\%, 99.9\%$ confidence level
respectively.

\subsection{Choice of binning } 
The absence of {\it an analytical form
of the luminosity function} has the drawback of preventing the use of a 2D
Kolmogorov--Smirnov test in the $(M_0,V/V_{max})$ plane.
So we are obliged to work in $n$ luminosity bins and an
arbitrary binning is substituted for an arbitrary analytic 
luminosity function. It is easier however to explore all the
possible binnings than to explore all the possible analytic expressions!
Moreover, the binning procedure is well suited to the method,
which applies whatever the choice of the bin $[M_0^1,M_0^2]$.

In principle, the choice of the binning is not constrained. However, 
two practical limitations arise:

$\bullet $ As noted by Lampton {\it et al} (1976), 
we should not proceed to
reject some values of the cosmological parameters from 
confidence levels without 
a preliminary good fit cosmological model. 
This is equivalent to requiring that the
lowest entropy ${\cal E}_{min}$ of the most probable cosmological model 
not exceed $n-3$ by much too much. This 
problem arises with too large $n$.

$\bullet $ The $x$ ratio is calculated from the fictitious 
redshift displacement
of a QSO according to the evolution law.
However it must not be forgotten that the evolution law has only a
statistical sense, and that it applies only to a QSO {\it population}.
If the number of QSOs per bin is too small (i.e. if $n$ is too large),
the statistical sense of the evolution law is lost and the
probability map becomes noisy. We choose a bin
number which keeps the smooth appearance of the probability map.

The method is validated by the analysis of synthetic
quasar catalogues which are described in Paper II.
By trial and error based on 
varying the size of the simulated QSO sample (see Section 5, Fig. 7), we 
found that the best choice from the point of view of the minimum entropy 
is to take approximately between 10 and 20 bins. 

Since a large part of the work is made with these catalogues, 
we discuss them in Section 5.

\section{How to compute $V/V_{ \, m\, a\, x \, }$ }

\subsection{Behaviour of $x$ versus the parameters}
In Paper II are introduced the various quantities necessary for
computing the individual $x$ ratios in a complete QSO sample. Two
curves in the Hubble
plane, the bright and the faint separators ${\cal S}_1$ and ${\cal
S}_2$, are of particular importance. These curves, shown in
Fig. 3, delimit the domain ${\cal D}$, lying between them, and the
domain ${\cal D}^*$ which is its complementary with respect to box
$\cal B$, or
$[m_1,m_2]\otimes [z_1,z_2]$. All quasars in domain ${\cal D}$ 
(here after $\cal D$ quasars) are redshift--limited, while all $\cal
D^*$ quasars are magnitude--limited.

According to Paper II, the $x$ ratio is uniformly distributed 
in the range $[0,1]$
for quasars in any absolute magnitude range $[M_0^1,M_0^2]$.
The $x$ ratio takes quite a simple form for $\cal D$ quasars:
\EQ
x \;=\; {V-V_1 \over V_2 - V_1 }
\EN
where the volumes $V_i=V(z_i)\;(i=1,2)$, the comoving volumes out to
redshift $z_i$, are determined only from the cosmological
model. For $\cal D^*$ quasars, on the contrary, there are 
many different cases according to the respective values 
of $z_1$ and $z_2$ and of the redshifts where the
magnitude equals one of the limiting magnitudes. An 
example is given in
Fig. 3. Consider the QSO denoted by $+$. If 
we exclude from the volume
computation the two redshift
 ranges where $m(z)<m_1$, then the $x$ ratio for this QSO is:
\EQ
x \;=\; {V-V_{min} \over V_{max}-V_{min} },
\EN
It should be noticed that $k_L$ does
not enter Eq. 6. $k_L$ concerns $\cal D$ quasars only inasmuch 
it governs the repartition of quasars 
with respect to the separators. 
On the contrary, $V_{min}$ and/or
$V_{max}$ which enter the equivalent of Eq. 6 for $\cal D^*$ 
quasars, do depend {\it explicitely} on $k_L$.

As evidenced in Fig. 2, the run of individual $x$ ratios
versus cosmological parameters varies substantially from quasar to
quasar, although $k_L$ is fitted to the maximum likelihood in each
cosmological model of Fig. 2. Large variations and even inversion of slope
correspond to the transition from one class to the other
(magnitude-- or redshift--limited). As $\Omega _{mat}$
increases at fixed $\Lambda $, and as $\Lambda $ decreases at 
fixed $\Omega _{mat}$, the curves tend towards horizontal
asymptotes so it becomes impossible to test between various
low $\Lambda $ or high $\Omega _{mat}$ universes. 
However we will introduce in the following an alternative method based on 
the synthetic catalogues analysis to avoid this problem.

\subsection{ Towards efficient cosmological tests }
Our method is far more demanding than a {\it
global application} of the $<V/V_{max}>$ test. 
We have several distinct sub-samples (the $N$ absolute magnitude
bins), which are independent in
the sense that they do {\it not} contain the same objects and their
$x$ ratios do not behave similarly, neither with respect to evolution 
strength, nor with respect to cosmological parameters.
The $x$-uniformity puts in $N$ constraints instead of a single one:
in some cosmological models, whatever the value of $k_L$, 
all the distributions of $x$ ratios have too low a probability to be uniform 
{\it simultaneously} for all the absolute magnitude bins.
This is exactly what is needed 
to reject some cosmological models as incompatible with a given
hypothesis on quasar evolution.

One shortcoming of this test is that the evolution hypothesis itself 
may be wrong, although it is quite widely used. However the values of
individual $x$ ratios for $\cal D$ quasars do not depend on the form and
the value of the luminosity evolution and only depend on the cosmology.
{\it If} there the proportion of $\cal D$ quasars is sufficient in current
samples then we will find some common results, even when varying
the various evolution hypotheses. We will see in Section 6 that this is
roughly the case.

There are various ways to perform the test:
the redshift range $[z_1,z_2]$ may be divided into bins. 
There is presumably an optimal way to operate this redshift binning:
since most of the effects of the cosmological constant 
are noticeable at high 
redshift, say $z>1$, the essential contribution of the range $[0.3,1]$ 
is likely to increase the noise. On the other hand the effects of
varying $\Omega _{mat}$ are already noticeable at lower redshift 
(Carroll {\it et al}, 1992), so that
one can expect to get constraints on $\Omega _{mat}$ from the low redshift
bin and constraints on $\Lambda $ from the high redshift bin.
There is probably an optimal compromise for the cutoff
to be found somewhere around $z=1$. In Section 6.1 we use this idea to
apply the test to low redshift ($0.3<z<1.5$) and high redshift ($1.5<z<2.2$)
samples of the Boyle QSO sample.

\section{ Monte--Carlo Simulations }

\subsection{Construction of Monte Carlo catalogues}
Before applying the test on real data, we checked its efficiency on
simulated samples. This
is the only way to vary the cosmological parameters, or to search for the
effect of the sample size, or of various biases in the redshift
distribution.

We generated Monte-Carlo samples of 400 objects for
each of the cosmological models of the grid
defined above, and we analysed the simulated samples exactly as we did
with Boyle's one. To simulate the catalogues, the procedures are the 
following: 

$\bullet $ choose a model (cosmology + evolution functional form)

$\bullet $ determine the function $\Phi (M_0)$
from the Boyle {\it et al.} global luminosity function.

$\bullet $ given a limiting magnitude and an absolute luminosity, 
compute the available volume $V_a(M_0)$

$\bullet $ sort in luminosity according to 
the observed probability distribution function 
of luminosities in a sample, that is the
product $\Phi (M_0) \times V_a(M_0)$ 

$\bullet $ given $V_a(M_0)$, sort in redshift ensuring homogeneous spatial
distribution.

The last two steps are repeated once per object in the sample.
All the details are described in Paper II.

\subsection{ Application of the test to simulated catalogues}

In this Section we give the results of the test on various simulated
catalogues by varying different parameters and by introducing some bias to
understand how the test works.
We choose to represent the contours at $90, 95, 99, 99.9 \%$ confidence
levels which are the minimum confidence levels to reject some models.
PWLE refers to power law luminosity evolution, and LEXP to exponential
luminosity evolution. The limiting magnitude of the simulated catalogues
is always $m_{lim}=21$. No bright limit is set, since it complicates 
the computations and does not change the test.

\subsection{ Comparing parent cosmologies }
Figs. 4 give the probability maps obtained in the
$(\Omega _{mat}, \Lambda )$ plane for 5 different parent cosmological 
models. All maps correspond to samples of 400 objects with the
same Monte--Carlo seed. All catalogues have been generated under
the PWLE assumption, nevertheless varying the functional form of the
evolution does not modify significantly the form of the contour maps.
A large part of the parameter space is rejected with the models at low
$\Omega_{mat}$. This is expected from the Fig.2 because the
stronger variations of the $x$ ratios are precisely for low $\Omega_{mat}$
cosmologies. Our test will be able to discriminate between low and
high $\Omega_{mat}$ values. A general characteristic of the contour maps
is the tendency to be drawn toward negative values of $\Lambda$. This too is
expected from Fig.2 where we note the asymptotic regime of the $x$
ratios toward these values. Our test is not able to discriminate between 
different values of $\Lambda$ with only 400 QSOs.

\subsection{ Binning} 
The effect of binning is shown in Figs. 5.1 to 5.6.
First, with a single bin, the test reduces to the global $<V/V_{max}>$,
and no model at all may be rejected. Only the models with no Big--Bang
(up--left of the map) are rejected because they are not computed!
As expected, increasing the binning decreases the allowed range in 
parameter space, until the
map becomes noisy and our method to find the
minimum entropy no longer applies.

\subsection{ The seed }
In order not to superimpose the effects of varying the seed and
cosmological effects, all the previous figures correspond to
Monte--Carlo catalogues drawn with the same initial seed.
To illustrate the effects of the statistical fluctuations, we show in
Figs. 6 the analyses of ten catalogues of the same
size and parent cosmology, which correspond to ten different
random seeds. The result is that varying the seed changes the form of 
the probability map, but does not reject the parent cosmological model.

\subsection{Sample size and limiting redshift}
Of course the larger the sample, the smaller is the probability map at
given confidence level and at fixed higher limiting redshift $z_2$.
This effect is illustrated in Figs. 7.1 to 7.4.
Despite the large number of QSOs in the
sample, the test is not able to determine $\Lambda$. The fact that the
test is not very sensitive to $\Lambda$ comes from the value of
$z_2=2.2$ which is not deep enough.

Conversely, the higher $z_2$, the more efficient is the test versus
cosmological parameters. The effect of increasing $z_2$ at fixed $n$
is shown on Figs. 8.1 to 8.3. 

In fact, these figures show that a high--z limited 
catalogue is very sensitive to
the curvature: on Fig.8.3 the contour probability
follows very well the iso--curvature line of the model, 
with only 400 QSOs in the sample!

To illustrate these effects, we give in Table 1 
the ranges corresponding to the 2
$\sigma $ (95 \%) confidence level for $\Omega _{mat}$ and $\Omega _k$
respectively.
However these results are quite doubtful beyond $z=2.2$ where 
we did not modify our models, while it
seems that evolution could reverse (see e.g. Hook {\it et al.} 1995).
Fortunately, the evolution governs only the maximum redshift of
magnitude--limited quasars, 
which are a minority (remember Section 4 and see Paper II).

\begin{table}
\caption{ 95 \% confidence level limits read in Figs. 7 and 8 
for $\Omega _{mat}$ and $\Omega _k$ for a parent
cosmology $\Omega _{mat}$ =0.6, $\Omega _k$ = -0.2 }
\halign{ \hfil # \hfill & \quad #\hfill & \quad #\hfill & \quad # &
\quad # \hfill \cr
\noalign{\hrule\medskip}
N    &$z_1$   &$z_2$& $\Omega _{mat}=0.60$ & $\Omega _{k}=-0.20$  \cr
&&&&\cr
400&0.3&2.2&$\Omega _{mat}=0.70\pm 0.50$&$\Omega _{k}=-0.15\pm 0.85$\cr
400&0.3&3.5&$\Omega _{mat}=0.70\pm 0.50$&$\Omega _{k}=-0.05\pm 0.55$\cr
400&0.3&5.0&$\Omega _{mat}=0.80\pm 0.6$&$\Omega _{k}=-0.05\pm 0.45$\cr
5000&0.3&2.2&$\Omega _{mat}=0.75\pm 0.35$&$\Omega _{k}=-0.1\pm 0.5$\cr
\noalign{\medskip\hrule}}
\end{table}

We combined the last two results to look for the most
efficient use of telescope time dedicated to cosmological tests based on 
quasars (see Section 7).

\begin{figure*}
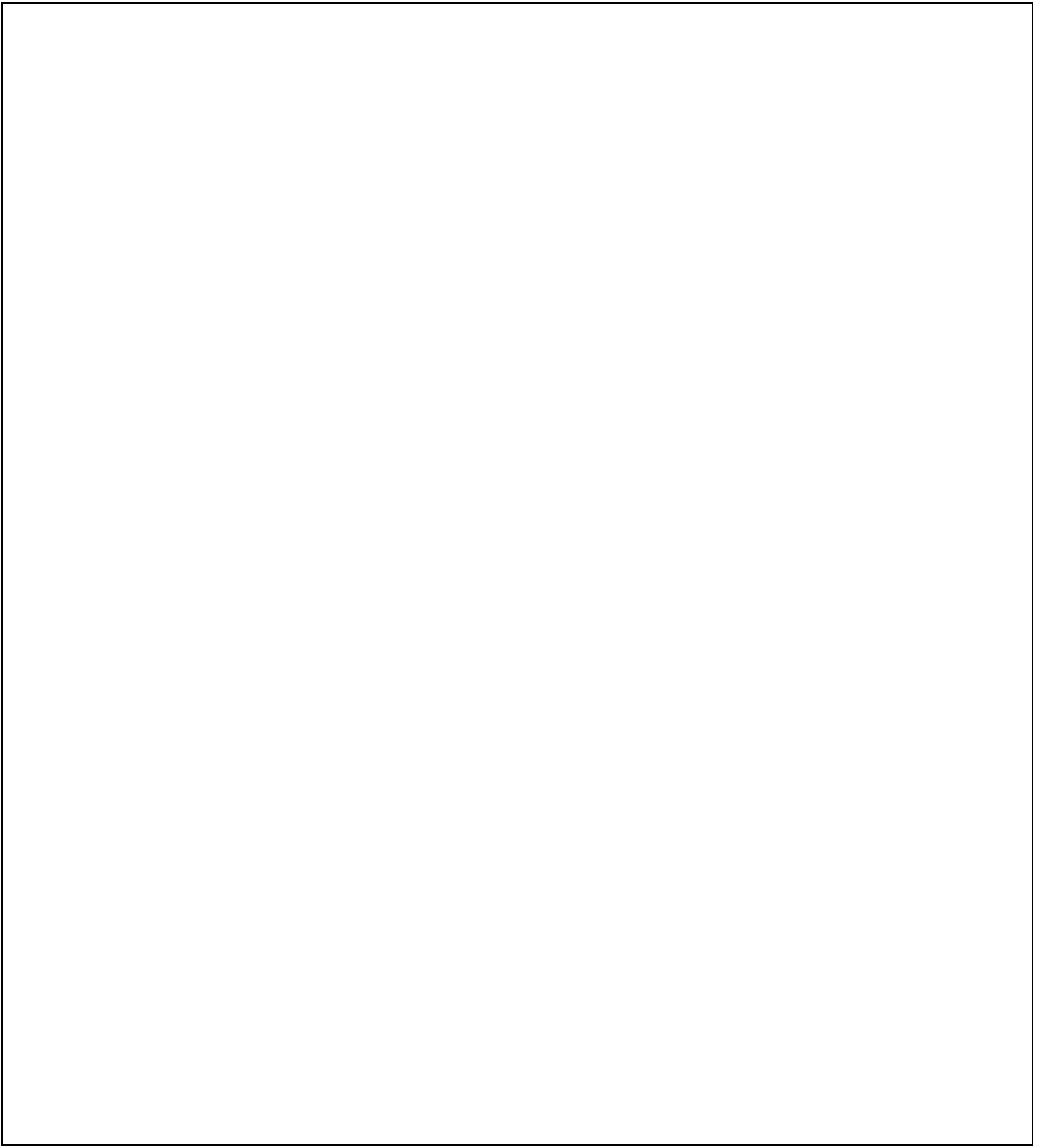

\picplace{20cm}
\caption{Contour probability maps varying the parent cosmological model of Monte--Carlo catalogues. Significance levels are 0.1, 1.0 5.0 and 10 \%.
All five models are constructed under PWLE with 400 QSOs and the redshift range [0.3,2.2]. They are analysed either under PWLE (left column) or under LEXP (right column). From top to bottom, the models are: ($\Omega_{mat}=0.0, \Lambda=0.0, k_L=2.95$); ($\Omega_{mat}=1.0, \Lambda=0.0, k_L=3.60$);($\Omega_{mat}=0.0, \Lambda=1.0, k_L=1.91$); ($\Omega_{mat}=1.0, \Lambda=1.0, k_L=3.82$);($\Omega_{mat}=2.5, \Lambda=1.0, k_L=4.12$). Note the strong difference between the probability maps for the low $(<0.2)$ and high ($>0.2)$ $\Omega_{mat}$.}
\end{figure*}
 
\begin{figure*}
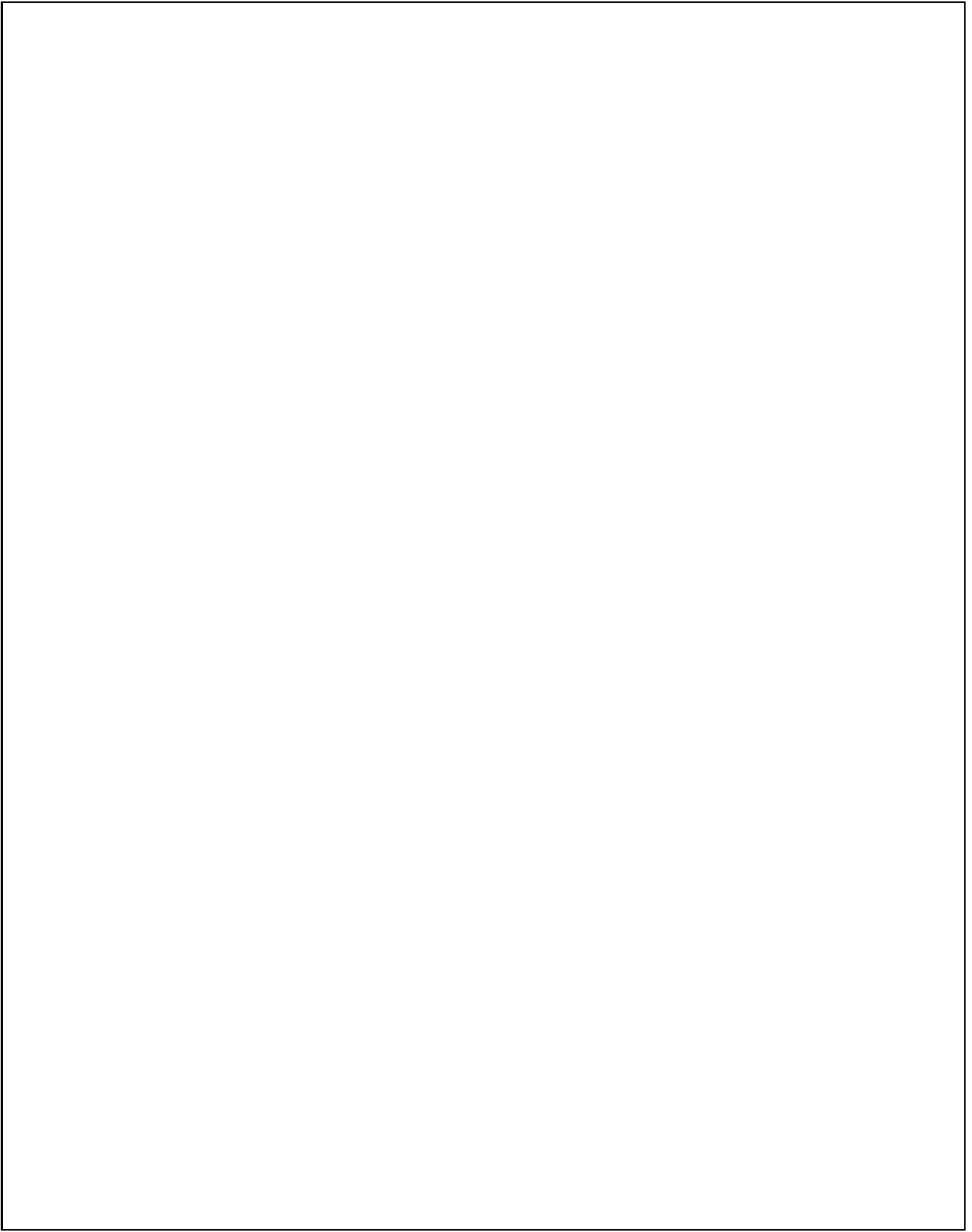

\picplace{23cm}
\caption{Same as Fig. 4, varying the binning. 
A PWLE simulated catalogue 
($\Omega_{mat}=\Lambda=0.6, k_L=3.47$) of 400 
QSOs, with a redshift range of [0.3,2.2] is analysed under PWLE hypothesis,
according to different binnings. From left to right and top to bottom,
the number of QSOs per bin is 400, 100, 50, 40, 27, 16.}
\end{figure*}
 
\begin{figure*}
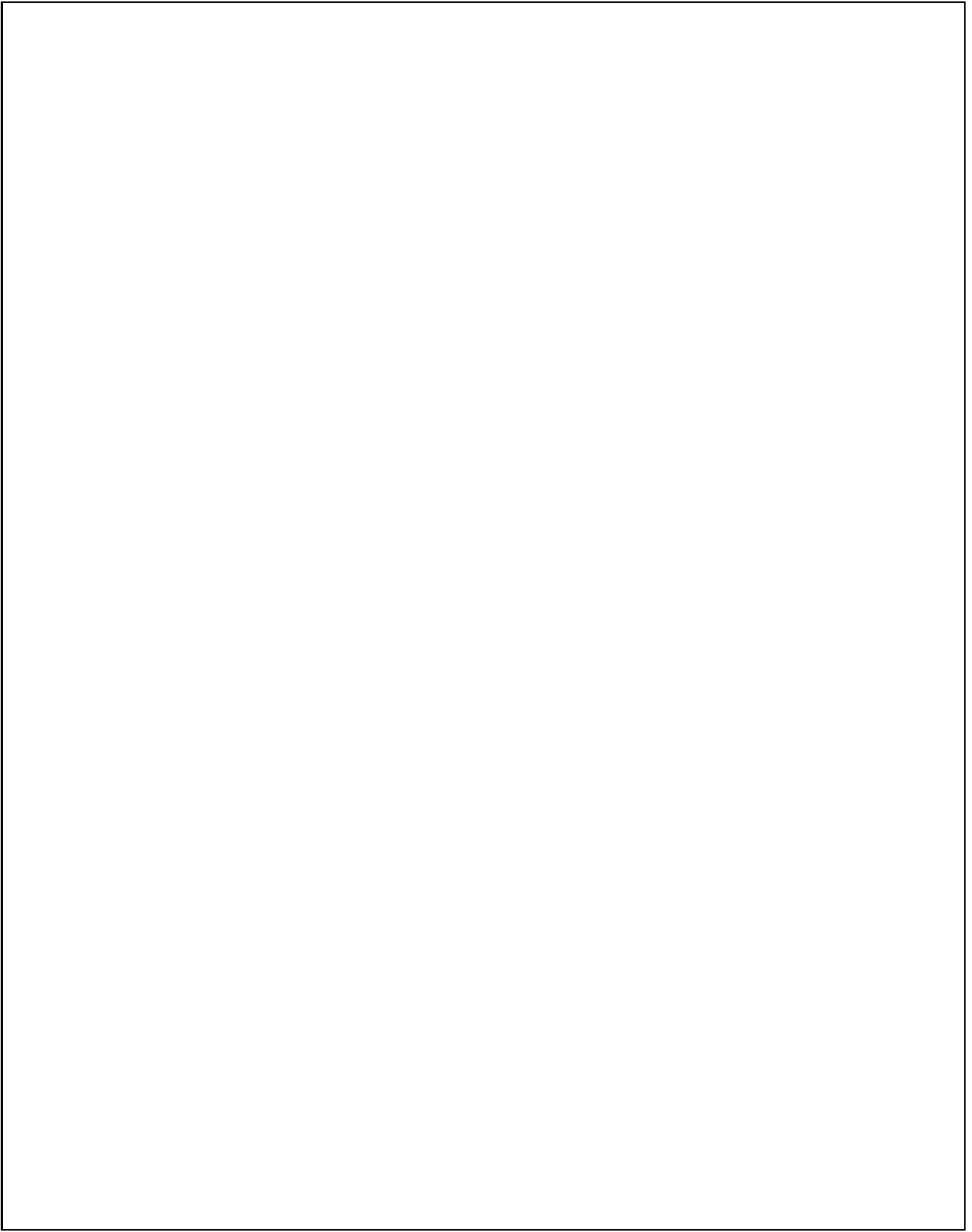

\picplace{23cm}
\caption{Same as Fig. 4, varying the seed. 
Ten different PWLE simulated 
catalogues in the
same cosmological model ($\Omega_{mat}=0.6 ,\; \Lambda=0.6,\; k_L=3.47$) 
are analysed under PWLE hypothesis.
All catalogues containe 400 QSOs in the redshift range [0.3,2.2], only the
random number seed is different in each catalogue.}
\end{figure*}

\begin{figure*}
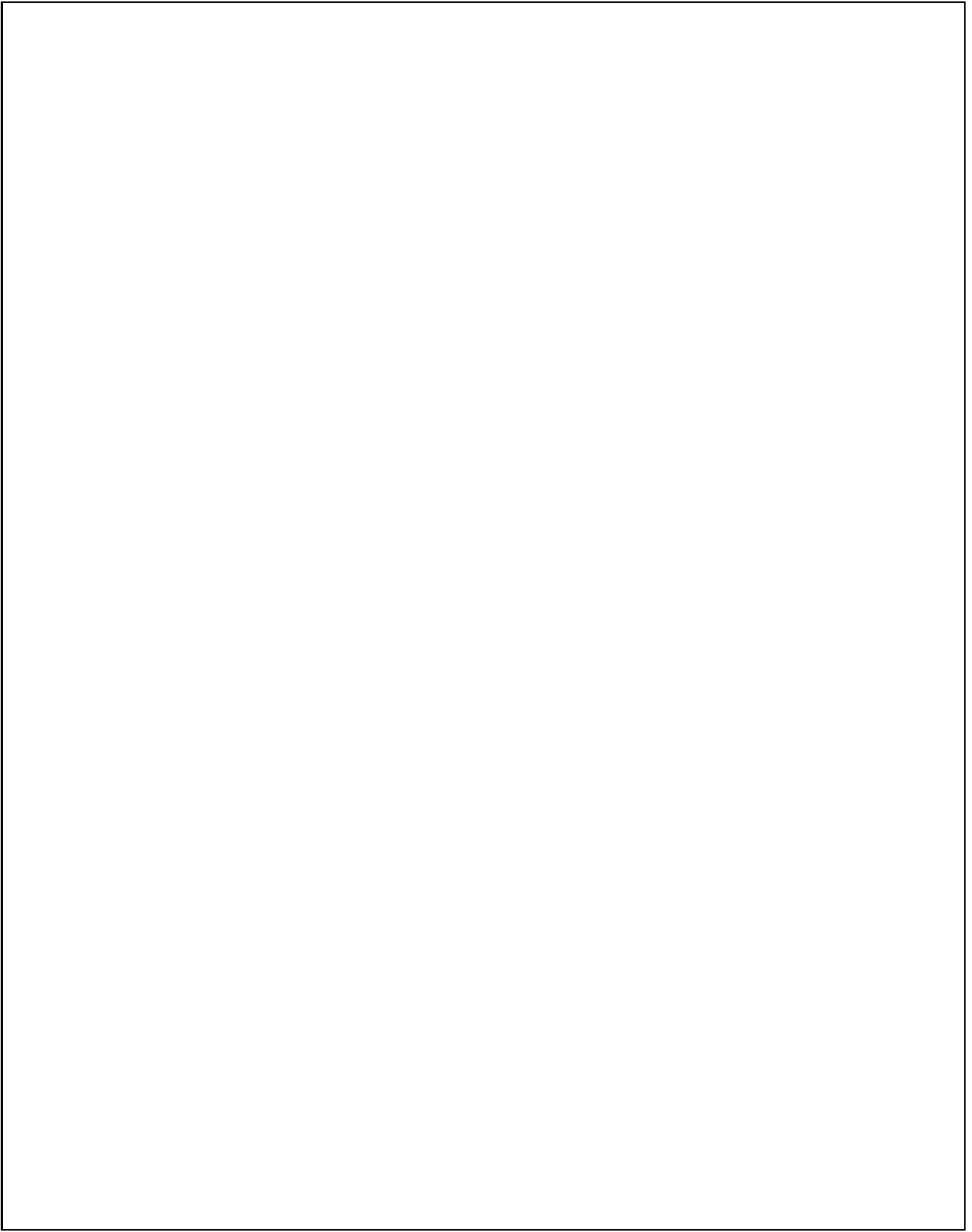

\picplace{23cm}
\caption{Same as Fig. 4, varying the catalogue size $n$.
Four PWLE simulated
catalogues in the same model ($\Omega=\Lambda=0.6, k_L=3.47$), the same
redshift range [0.3,2.2] are analysed under PWLE hypothesis. The values of 
$n$ are 400,1000,3000,5000 QSOs from top to bottom and from left to
right.}
\end{figure*}
 
\begin{figure*}
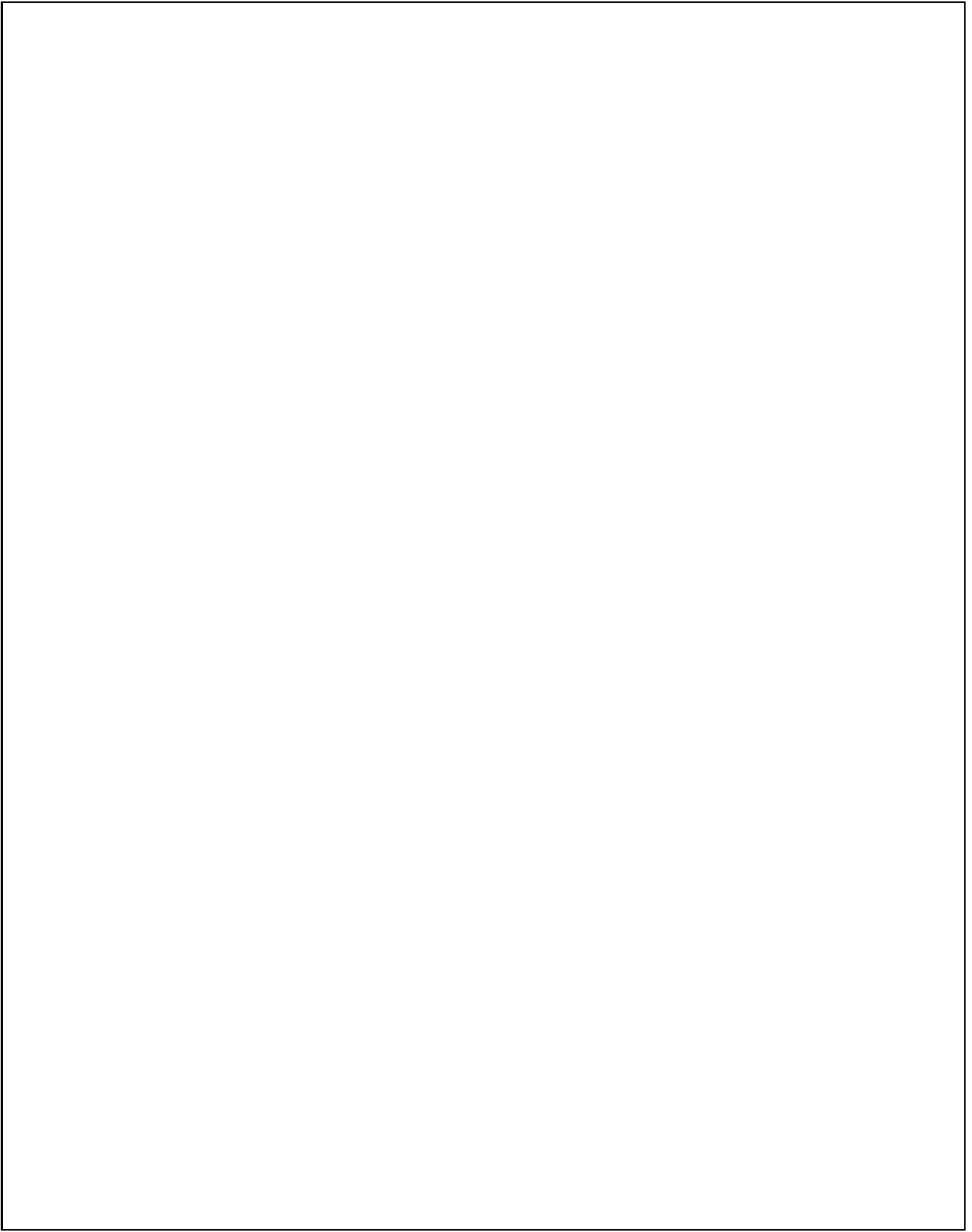

\picplace{23cm}
\caption{Same as Fig. 4, varying the upper redshift limitation
of the catalogue. Three PWLE
simulated catalogues with 400 QSOs, in the same model
($\Omega=\Lambda=0.6, k_L=3.47$) are analysed under PWLE hypothesis. The
lower redshift is always $z_1=0.3$, and the three values of the upper
redshift are $z_2=2.2, 3.5, 5.0$.}
\end{figure*}

\begin{figure*}
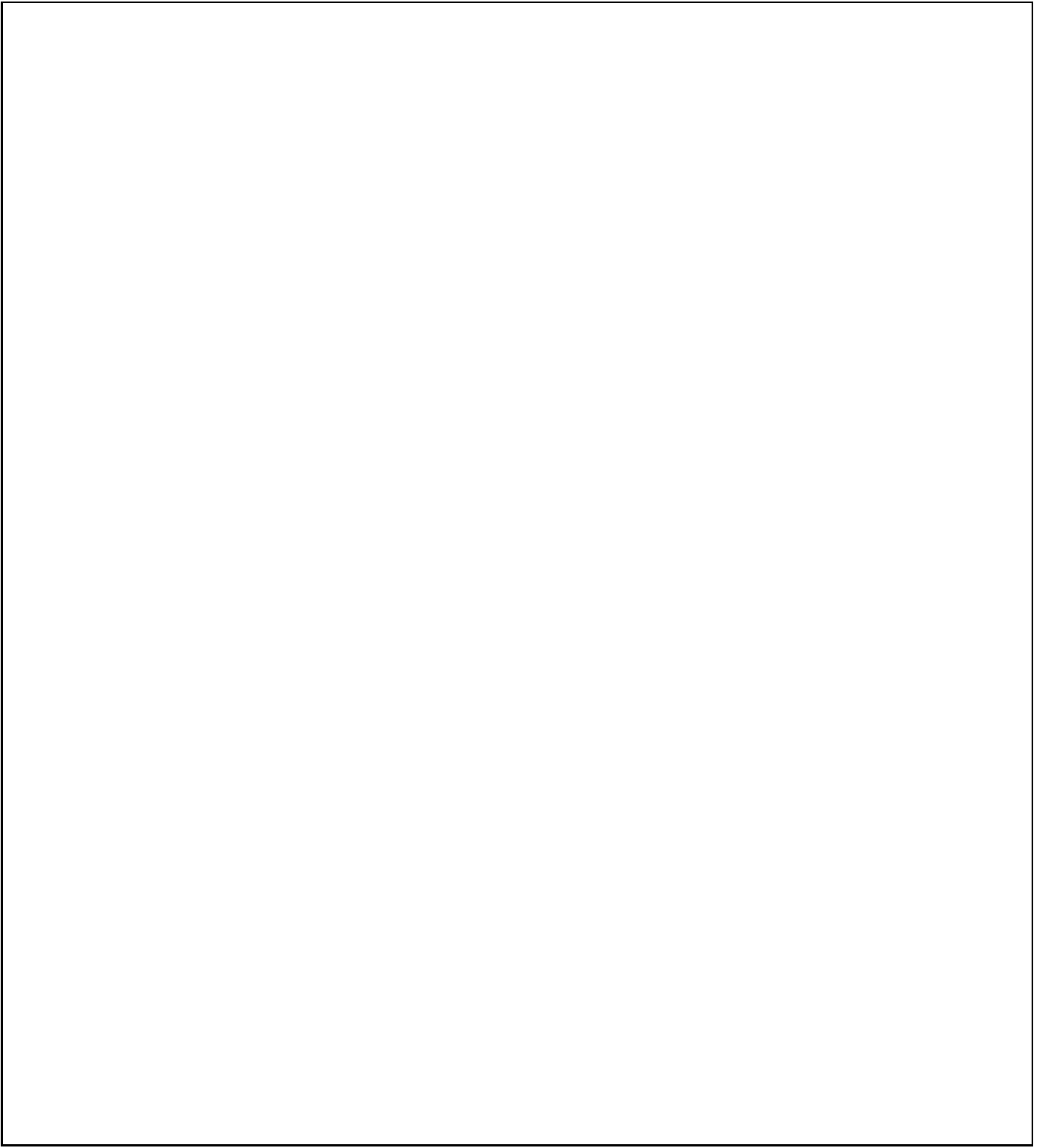

\picplace{20cm}
\caption{Effects of the redshift and magnitude biases. The unbiased
catalogue contains 546 QSOs, PWLE, in the model $\Omega=\Lambda=0.6,
k_L=3.47$, and the redshift range [0.3,2.2]. All analyses are done under
PWLE hypothesis. The first
map (Fig. 9.1) is the analyse of the unbiased simulated catalog.
In the top right map, $30 \%$ of the QSOs in the redshift range [0.5,0.9]
are subtracted. In the middle left map $30 \%$ of the QSOs in the redshift
range [1.5,1.8] are subtracted. In the fourth map the last two
redshift biases are present. In the fifth map only a gaussian noise in
magnitude, of mean 0.0
and deviation 0.01, and a second component of mean 0.2 and deviation 0.2
has been done. The last map contains all the previous biases.}
\end{figure*}

\subsection{Biasing }
\subsubsection{Biasing due to lensing} 
At given redshift,
a presumably small, but still unknown, fraction of QSOs 
are amplified while the other ones are deamplified. 
The probability that a (lens) galaxy lies closer to the line-of-sight 
of a given quasar than
some threshold sufficient to induce gravitational 
amplification increases with increasing quasar redshift. 
As amplified quasars enter preferentially magnitude-limited
samples, the result of lensing is to mimic intrinsic luminosity
evolution. Unfortunately, we do not know the respective contributions of
intrinsic and extrinsic biasing to the observed evolution. As 
the evolution law
we use is purely phenomenological, and not the prediction 
of e.g. a physical model of the 
emission in the environment of a massive black hole, it can fit
the effects of lensing as well - or as badly - as those of
an intrinsic luminosity evolution.

\subsubsection{Clustering} 
Our test relies on the null hypothesis of
uniform spatial distribution, which is in principle not
compatible with clustering, if any. In fact it 
is negligible here since our working scale is of the order
of the Gpc, and only a weak clustering scale of 40-100~Mpc seems to be
detected (Clowes and Campusano 1991, Shanks {\it et al}, 1994, 
Georgantopoulos {\it et al}, 1994).

\subsubsection{Biasing in $z$}
In order to be as realistic as possible, 
samples with a deficiency of objects in various redshift
ranges have been used. According to Boyle {\it et al.} 
(1987; see also
V\'eron 1983), the combination of the run of color versus 
redshift with a pre-selection based on UV excess results in 
a defficiency around 30\% of objects in the redshift 
range $0.5<z<0.9$. We analysed samples with such a random
depletion in this range. We also studied the effects of a
similar defficiency in the range $1.5<z<1.8$.
Figs. 9.2, 9.4, and 9.6 show the effect of this defficiency, which clearly
results in a general enlargement of the probability map. Fortunately we have
not detected a strong shift of the contour probabilities. 

\subsubsection{Biasing in magnitude}
The photometric biasing has been analysed. We introduced a gaussian
noise of mean 0.00 and standard deviation 0.01 mag. To mimic QSO
variability, a second component of mean 0.2 and standard
deviation 0.2 has been added (Giallongo {\it et al.} 1991).
In Fig. 9.3 there is no redshift bias, but the two photometric 
biases are included, and Fig. 9.5 has
been drawn with the two redshift biases plus the two photometric biases. 
The absence of strong differences
with the non--biased map (Fig. 9.1) shows that the $x$ ratios of the
$\cal D$ QSOs do not depend on their magnitude. 
Fig. 9.5 shows a map enlarged from Fig.9.1, which is the expected
effect caused by any bias.

\section{Real QSO Samples}

\subsection{ Application to real quasars}
We now apply the test to the sample of 400 UVX quasars
of Boyle {\it et al.} (1990). This sample is based on 
fiber FOCAP spectroscopy
of 1409 UVX objects selected through COSMOS photometry 
in eight high-latitude Schmidt fields. The limiting $B$ magnitudes and
$U-B$ color indices are given in Boyle {\it et al.}'s Table 2 (1990).

Boyle {\it et al.} (1988) show that the best fit to the data in this
sample is
obtained with Pure Luminosity Evolution (PLE) models, far better than 
PDE. The favoured functional form for LE seems to
be a power law (PWLE) rather than an exponential (LEXP).

Both evolution models depend on a single parameter $k_L$ (see Section
2). Fig. 10 gives the probability map obtained with PWLE and LEXP, which
appear to be essentially similar.
Restricting ourselves to the $\Lambda >0$ half plane, we discuss the
results at the 95 \% confidence level:

1) Regarding the matter term, we find, from Fig. 10.a,b that
$0.3<\Omega _{mat}<1.3$ (PWLE) and $0.2<\Omega _{mat}<2.$ (LEXP).
The minimum entropy in both models is the same and the corresponding
probability is $52\%$. More interestingly,
the two functional forms (power law and exponential) 
lead to similar probability maps, which was expected from the analysis
of the simulated catalogues.

2) In Fig. 10.c,d,e,f the two probability maps obtained at low-
($0.3<z<1.5$) and high- ($1.5<z<2.2$) redshift respectively are compared.
The latter range does not allow any rejection among
cosmological models, although the 'cosmological signal' is believed
to increase with redshift. This is due to the
narrowness of the redshift range which does not allow one to discriminate
cosmologies, and to the poor sample size (only
112 high--z QSOs). The low-redshift map leads 
to the following constraint: $\Omega_{mat}< 0.7 $ (PWLE) and 
$\Omega_{mat}< 0.9$ (LEXP). As expected from simulated catalogues, at
low redshift range, the test is not sensitive to $\Lambda$.

3) Concerning the space curvature, the limits are the following. With
QSOs over the entire redshift range $[0.3,2.2]$ we read  from Fig. 10.a,b:
$-1.1<\Omega_k <0.7 $ (PWLE) and $-1.1<\Omega_k <0.8$ (LEXP).
In the redshift range $[0.3,1.5]$ we find from Fig. 10.c,d,e,f:
$-0.7<\Omega_k <1.$ (PWLE), and $-0.8<\Omega_k <1.$.

The constraint on the cosmological constant is only $\Lambda < 1.4$ for
both evolution models.

It does not make sense to multiply the probability map of the
low--z sample with that obtained from the entire sample because the
samples are not independent. However we can restrict the space parameter
to the intersection of the space parameter given from these two samples. 
The results are summarized in Table 2. The $1 \sigma$ confidence levels
are given in Table 3.

\begin{table}
\caption{ 95 \% confidence limits on $\Omega _{mat}$, $\Omega _k$, and
$q_0$ from low--z and all QSOs together}
\halign{ \hfil # \hfill & \qquad \hfill #\hfill &\qquad \hfill
#\hfill \cr 
\noalign{\hrule\medskip}
\multispan3 \hfill $0<\Lambda <1.4$ \hfill \cr
\noalign{\medskip}
\noalign{\hrule\medskip}
&PWLE&LEXP\cr
$\Omega _{mat}$ &$0.5 \pm 0.2$&$0.55 \pm 0.35$\cr
$\Omega _k$ &$0.0 \pm 0.7$&$0.0 \pm 0.8$\cr
$q_0$ &$-0.25 \pm 1.0$&$-0.2 \pm 1.3$\cr
\noalign{\medskip\hrule}}
\end{table}

The Luminosity Functions derived from the sample 
and the estimation of the characteristic evolution times in best
models are given in Paper II.

\begin{table}
\caption{ 68 \% confidence limits on $\Omega _{mat}$, $\Omega _k$, and
$q_0$ from low--z and all QSOs separately}
\halign{ \hfil # \hfill & \qquad \hfill #\hfill &\qquad \hfill
#\hfill \cr 
\noalign{\hrule\medskip}
\multispan3 \hfill $0.3< z <2.2$ \hfill \cr
\noalign{\medskip}
\noalign{\hrule\medskip}
&PWLE&LEXP\cr
$\Omega _{mat}$ &$0.75 \pm 0.35$&$0.55 \pm 0.25$\cr
$\Omega _k$ &$-0.05 \pm 0.65$&$-0.15 \pm 0.55$\cr
$q_0$ &$0.1 \pm 1.2$&$-0.3 \pm 0.9$\cr
\noalign{\hrule\medskip}
\multispan3 \hfill $0.3< z <1.5$ \hfill \cr
\noalign{\medskip}
\noalign{\hrule\medskip}
&PWLE&LEXP\cr
$\Omega _{mat}$ &$0.2 \pm 0.2$&$0.25 \pm 0.25$\cr
$\Omega _k$ &$0.35 \pm 0.65$&$0.25 \pm 0.75$\cr
$q_0$ &$-0.35 \pm 1.0$&$-0.4 \pm 1.1$\cr
\noalign{\medskip\hrule}}
\end{table}

\begin{figure*}
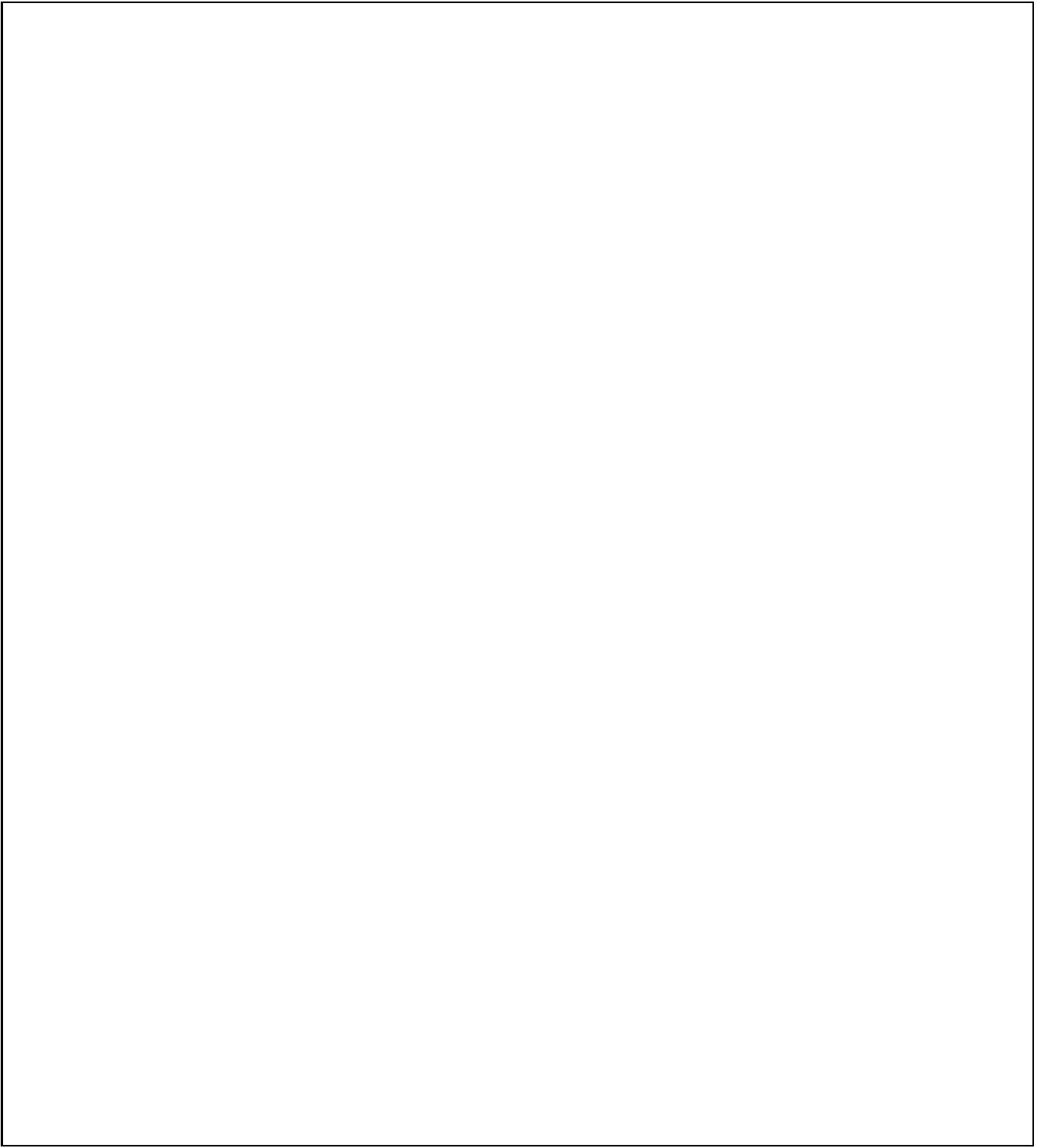

\picplace{20cm}
\caption{ Contour probability maps (0.1, 1.0, 5.0 
and 10 \%
significance level from outside to center) of the Boyle catalog. Left
(right) column is the PWLE (LEXP) analysis. At the top:
the whole redshift range $[0.3,2.2]$, in the middle:
the restricted low--redshift range $[0.3,1.5]$, and in the bottom: the
high--redshift range $[1.5,2.2]$. The latter range is too small to give
useful limitations.}
\end{figure*}

\section{Summary and Discussion}
\subsection{ Summary }
There are several points to recall before discussing the results found
in Sections 5 and 6. 

All the numerical codes we have designed for the present work
have been duplicated, independently by two of us, in order to avoid 
computational errors.

None of the steps of the entire test procedure 
depends on the Hubble constant $H_0$.
The test is completely independent of any absolute magnitude or
distance scaling, 
since it only relies on volume ratios and on flux ratios. 
So, the only parameters which do depend on the choice of $H_0$ are
the characteristic evolution time and the quasar absolute
luminosities such as $\Phi_{*}$, the normalization of the 
luminosity function given in Paper II. 
Schade and Hartwick 1994, on the contrary, need an explicit 
value of $\Phi_{*}$, thus a choice of $H_0$. They also make use 
of the same functional form and
parameters for the luminosity function in all cosmological models. Even
if the values used are within error bars for all models, they
are better suited to one or more peculiar model, which is 
likely to bias this test.

Our results on the cosmological parameters only apply {\it under
specific hypotheses concerning quasar evolution}. 
Such hypotheses are difficult to avoid and the hypotheses made
here are quite widely encountered. The present paper is
entirely devoted to the case for Luminosity Evolution, which has been 
favoured by recent studies (Boyle et al. 1988, 1990). 
There is a way to test the adopted evolution laws: applying the
$<V/V_{max}>$ test in various redshift bins, and comparing the values
found for $k_L$. Such a procedure however requires quite a large
sample. Applying the test in the frame of Density Evolution 
will be the subject of a forthcoming paper. 
Quite interesting are the results of cross-tests of the two
standard functional forms: samples issued from a simulation 
with power law (PWLE) and tested with exponential (LEXP)
lead to similar probability maps in the
$(\Omega _{mat},\Lambda )$ plane, which makes
us rather confident on their generality. Hopefully, such similarities 
may be expected from cross-tests between Density and Luminosity
Evolutions, which will follow in a forthcoming paper.
The test seems to be insensitive to the cosmological constant. In fact
we showed it is sensitive to a certain function $f(\Omega_{mat},
\Lambda)$ which depends on the redshift depth of the catalogue.
$f(\Omega_{mat},\Lambda)=\Omega_{mat}$ for a catalogue with a low  
redshift limitation $(z_2 \simeq 2.2)$. $f(\Omega_{mat},\Lambda)$
tends to the curvature $\Omega_{mat}+\Lambda$ for a deep catalogue. 
By this way, with at least two different catalogues, it is possible to 
determine $\Lambda$. The recent LBQS catalogue 
of Hewett {\it et al} (1995), with more than 1000 QSOs, 
may be useful for this purpose.

\subsection{ Limits on $\Omega _{mat}$, and the status of $\Lambda$}
As already underlined, the constraints brought in by the cosmological
test described in this paper hold mainly on $\Omega _{mat}$, and
to a lesser extent on the curvature term $\Omega _k$; see Table 2.
A low $\Omega _{mat}$ is rejected, and we exclude a flat Universe
without a cosmological constant. This implies introducing a non zero
$\Lambda$ in the standard inflation models ($\Omega_{tot}=1$).
This is in the sense of the recent determinations of the Hubble
parameter which give too low an age for the Universe if $\Lambda=0$, 
compared to the age of the oldest stars only (Jaffe, 1995).

Let us compare our limits on $\Omega _{mat}$ to what is found 
in the current litterature, arising from tests of different nature
and/or on different scales:
The limits coming from velocity fields in voids are
obtained without assumptions regarding either the galaxy biasing, $\Lambda $, 
or the fluctuation statistics (Dekel and Rees 1994). 

White {\it et al.} (1993) express $\Omega _{mat}$ in terms of 
$f_b$, the baryon fraction in clusters of galaxies: 
$\Omega _{mat}< 0.3 \left({\Omega _b \over 0.06} \right) \left({0.2 \over
f_b } \right) $, {\it if} the ratio of baryonic to total matter in
clusters is representative of the universe.
Accounting for the limits on
the baryon fraction deduced from 
X-Ray observations of the Coma cluster, 
$ f_b = 0.009 + 0.05 h^{-3/2} $, and on the limits on 
$\Omega _b$ from standard Big-Bang Nucleosynthesis: 
$0.009 h^{-2}< \Omega _{ baryon}< 0.02 h^{-2}$, 
we finally get: 
$\Omega _{mat}< 0.4 {h^{-1/2} \over 1+ 0.18 h^{3/2}}$.
Comparing our results with two recent papers is of particular interest: 
Bernardeau {\it et al.} (1995) fix $\Omega >0.3$ 
at the 2 $\sigma $ level and favour $\Omega =1$ from the skewness of the
cosmic velocity field divergence. Meanwhile 
Hudson {\it et al.} (1995), by comparing the distribution of optical
galaxies with the POTENT mass distribution, conclude that 
$\beta =\Omega ^{0.6}/b=0.74 \pm 0.13 \; (1 \sigma ) $, 
favouring either 
$\Omega =1, \; b=1.35 \pm 0.23 $ or $b=1, \; \Omega =0.6 \pm 0.18$. 
Cole {\it et al.} (1995) describe similar work from IRAS
redshift surveys and find that $\beta =0.5 \pm 0.2 $,
implying
$\Omega =1, \; b\simeq 2 $ or $b=1, \; \Omega =0.35 $.
This last two results are
surprinsingly close to ours: $ \Omega =0.5 \pm 0.2 \; (2 \, \sigma)$
(PWLE) or $ \Omega =0.55 \pm 0.3 \; (2 \, \sigma)$ (LEXP).
Perhaps we can conclude that the bias parameter $b$ is of order unity,
but this still appears to be premature!

All these results are summarized in Table 4.
\begin{table*}
\caption{  Limits on $\Omega _{mat}$ obtained in various recent
tests }
\halign{ \hfil # \hfil & \hfil #\hfil &\hfil
#\hfil &\hfil #\hfil & \quad \hfil # \hfill\quad \cr
\noalign{\hrule\medskip}
\multispan5 {\it (Sp. H. = spiral haloes, CoG=Clusters
of Galaxies, $f_b$=baryon fraction,  VT=Virial 
Theorem, CVT=Cosmic VT)} \cr
\noalign{\medskip}
\noalign{\hrule\medskip}
&&&&\cr
Test & $\Omega $& confidence level & $\Lambda $& reference \cr
&&&&\cr
Galaxies:&&&&\cr
&&&&\cr
Sp - H. & $0.1 < \Omega < 1$&$1.5 \sigma $ & $\Lambda =$ 0 &
 Zaritsky and White (1993)\cr
&&&&\cr
\noalign{\smallskip\hrule\smallskip}
Clusters:&&&&\cr
&&&&\cr
CoG: VT &$ \Omega h_{75}^{-1} = 0.2 \pm 0.1 $ && $\forall \Lambda $&\cr
CoG: Giant Arcs & $ \Omega h_{75}^{-1}= 0.2 \pm 0.1 $ &&$\Lambda $=0& Fort and Mellier (1994) \cr
CoG: Weak Shear &$ \Omega h_{75}^{-1}= 1 \pm 0.5 $ &&$\Lambda $=0& Bonnet {\it et al. } (1994) \cr
&&&& Fahlman {\it et al. } (1994) \cr
CoG: $f_b$ &$\Omega h_{75}^{1/2} <0.5 $ &&$\Lambda $=0& White {\it et al. } (1993) \cr
\noalign{\smallskip\hrule\smallskip}
&&&&\cr
Velocity Fields:&&&&\cr
&&&&\cr
CVT-CfA1 & $ \Omega = 0.5 \pm 0.1 $ &&$\forall\Lambda $&\cr
POTENT--IPDF & $0.3 < \Omega $&$5 \sigma  $ &$\forall\Lambda $& Dekel (1994)\cr
POTENT--VOIDS &0.3 $< \Omega $ & $ 2.4 \sigma $ &$\forall\Lambda $& 
Dekel and Rees (1993) \cr
POTENT--SKEW & 0.3 $< \Omega $ & $ 2\sigma  $ & $ \forall\Lambda $ & Bernardeau {\it
et al. } (1995) \cr
\noalign{\smallskip\hrule\smallskip}
&&&&\cr
QSOs:&&&&\cr
&&&&\cr
spatial distribution& $ \Omega = 0.4 \pm $0.2 &1 $ \sigma $ & 
$ \Lambda =$ 0 & Deng et al. (1994)\cr
Loh--Spillar & $\Omega = 0.2\pm $ 0.1 & 1 $\sigma $ & $ \Lambda $ =0&\cr
& $ \Omega = 0.4\pm 0.1 $ &1 $ \sigma $ & $ \Lambda  =1-\Omega $ & Schade 
et al. (1994)\cr
This Work: & $ \Omega =  0.5 \pm 0.2 $ & $ 2\sigma $ &
$ \Lambda<$ 1.4 & PWLE\cr
&  $ \Omega = 0.5 \pm 0.3 $ & $2\sigma $ & $ \Lambda<1.4 $  & LEXP \cr
&&&&\cr
\noalign{\medskip\hrule}}
\end{table*}

The test described in the present paper, performed with QSOs, 
measures the cosmological 
parameters on a scale of several $Gpc$. It agrees well 
with most of the tests given in Table 4. The only exception
concerns the results on the centers of clusters of galaxies (both 
the Virial Theorem and the Giant Arcs) which
have been operated on smaller scales, typical of structures which have
detached from the general expansion, then collapsed and dissipated. 
Furthermore, these
tests rely on the assumption of a universal Mass to Light
ratio, and it is not surprising
to find smaller $\Omega _{mat}$ on smaller scales.




\bigskip




 
\subsection{ Optimal strategy for quasar detection aimed at cosmology }
We combined the results of the last two series of simulations to look 
for the most efficient use of telescope time aimed at cosmological 
tests based on quasars. 
We saw that for the same precision on the cosmological parameters, say 
$\Omega _{tot}$, about 10 times less QSOs are needed if their limiting
redshift is 3.5 instead of 2.2.
Let us compare two possible observational strategies for quasar
detection, for example with
the FUEGOS multifiber spectrograph to be build for the VLT at ESO:
first, programme 1, a classical preselection of UVX candidates brighter
than $B=24$ with some limit in $U-B$, and second, programme 2, 
consisting off the
spectroscopy of all stellar or compact objects brighter than $R=23$ (
Mathez {\it et al}, 1995a).
A selection of $I<22.5$ could be
preferred.
A preselection of UVX candidates (or simply through the blue 
filter, see Glazebrook {\it et al.} 1995) will be efficient 
with a QSO/candidates ratio around 1/3. The drawback, however, is
the redshift limitation inherent in the blue selection. 
In programme 2, the drawback is that the $QSO/candidate \, \,
ratio$ is far (to what extent ?) lower.
The question is wether this drawback may 
be compensated by the higher
limiting redshift, which would be higher than 2.2. 
The last source of error is that we used a simulated catalogue with an
evolution extrapolated from low redshifts to $z=3.5$. 
Although the evolution law at these
high redshifts is unknown, it is probably different from the evolution at low
redshift.
The two strategies are compared in Table 4, note however 
that many input quantities in Table 4 are only first guesses.
The surface density $\Sigma =300 \, deg.^{-2}$ of QSOs 
brighter than $B=24$ is one of the most
secure inputs (to within a factor of 2) since convergent 
results have been found by
Glazebrook {\it et al.} (1995) and Koo, Kron and Cudworth (1986).
The QSO/candidate ratio expected in programme 2 was taken as 1/30
to compare the two programmes. We think that it is rather conservative
for a selection through the red filter and is likely to allow a higher
limiting redshift. The quasar evolution at redshifts above 2.5 is
also questionable. It seems to reverse (Hook {\it et al.} 1995),
but it is not easy to ascertain because of the different observational
approaches below and above this redshift.
The surface density of stars to 23rd magnitude has
been estimated from the model of Robin (1989). As for the surface density of
quasars to $R=23$, Table 4 gives a conservative estimate since no
quasar selection has ever been made in $R$. Recent results by Webster
(1994) seem to indicate that red quasars are not rare, and that 
the surface density of red or infrared QSOS could be far higher. 

With the 80 fibres of FUEGOS, both programmes may be expected to have
similar
global efficiencies for the determination of $\Omega _{mat}$.
However they would lead to different scientific outputs: a
complete sample of 10,000 QSOs in the first case, allowing excellent
statistics for quasar studies, and a better
idea of the QSO evolution at high redshift, i.e. presumably constraints
on galaxy formation, in the second one.
And a good model of high--z evolution is what we need 
to determine $\Lambda$ from this test.

\begin{table}
\caption{ Optimal strategy for detection of quasars with FUEGOS with
80 fibres (FF= FUEGOS FIELD = 1/7 sq. deg.)}
\halign{ \hfil # \hfill & \quad #\hfill & \quad #\hfill & \quad #
\hfill \cr
\noalign{\hrule\medskip}
& $U-B<0.2$ & No Preselection \cr
$m_{lim}$ & B=24 & R=23 \cr
$\Sigma _{cand}\,(deg^{-2}$)&   1050  & 7500  \cr
$\Sigma _{QSO} \,(deg^{-2}$)&    300   &  250  \cr
QSO/candidate &    1/3.5   &  1/30  \cr
candidate/FF         &    150 &  1100  \cr
QSO/FF        &  50 &   35 \cr
exposures/FF \, (80 \, fb)  & 2 &  14  \cr
$N_{QSO} \, (\sigma _{\Omega _k}=0.5)$ & 5000& 400\cr
survey area $(deg^2)$ & 17& 1.6 \cr
FF & 100 & 12 \cr
total exposures & 200 & 170 \cr
\noalign{\medskip\hrule}}
\end{table}

\paragraph{Acknowledgements}
It is a pleasure to acknowledge B. Fort, P.Y. Longaretti, 
S. Andreon, J. Bartlett, J.F. Le Borgne, L. Nottale 
R. Pell\'o, JP Picat, G. Soucail, S. Collin--Souffrin for 
all the enlighting discussions we had on QSOs and cosmology, 
and especially T. Bridge for a careful reading of the manuscript.
L.V.W. thanks the french MESR for grant 93135.
This work was supported by grants from the french CNRS (GdR Cosmologie), 
from the
European Community (Human Capital and Mobility ERBCHRXCT920001).

\section{References}
\parindent 0truemm
Alcock, C., Paczynski B., 1979, MN 281, 358
\vskip 1truemm

Avni, Y., Bahcall J., 1980, ApJ 235, 694
\vskip 1truemm

Bernardeau, F., Juszkiewicz, R., Dekel, A., Bouchet, F., 1995, \par
\hskip 2truemm MNRAS 274, 20
\vskip 1truemm

Bigot, G., Triay, R., 1991, Phys. Let. A 159, 201
\vskip 1truemm

Bonnet, H., Mellier, Y., Fort, B., ApJ 427, L83
\vskip 1truemm

Boyle, B.J., Fong, R., Shanks, T., Peterson B.A., 

\hskip 2truemm 1987, MN 227, 717
\vskip 1truemm

Boyle, B.J., Shanks, T., Peterson B.A., 1988, MN 235, 935
\vskip 1truemm

Boyle, B.J., Fong, R., Shanks, T., Peterson B.A., 

\hskip 2truemm 1990, MN 243, 1
\vskip 1truemm

Broadhurst, T., Ellis, R., Koo, D., Szalay, A., 1990, 

\hskip 2truemm Nature 243, 726
\vskip 1truemm

Carroll, S.M., Press, W.H., Turner, E.L., \par
\hskip 2truemm 1992, Ann. Rev. of A\& A 30, 499

Crawford, D.F., 1995, ApJ 441, 488
\vskip 1truemm

Dekel, A., 1994, Ann. Rev. A. \& A 32, 371
\vskip 1truemm

Dekel, A., Rees, M., 1994, ApJ 422, L1
\vskip 1truemm

Deng, Z., Xia, X., Fang, L.Z., ApJ 431, 506
\vskip 1truemm

Efron, B., 1979, Ann. Statist. 7, 1
\vskip 1truemm

Fahlman, G.G., Kaiser, N., Squires, G., Woods, D., \par
\hskip 2truemm 1994 ApJ 437, 56
\vskip 1truemm

Fort, B., Mellier, Y., 1994 A\& A Rev. 5, 239
\vskip 1truemm

Freedman, W.L., Madore, B.F., Mould, J.R., Hill, R.,

\hskip 2truemm Ferrarese, L., Kennicutt Jr. R.C., 

\hskip 2truemm Saha, A., Stetson, P.B., Graham, J.A.,

\hskip 2truemm Ford, H., Hoessel, J.G., Huchra, J.,

\hskip 2truemm Hughes, S.M., Illingworth G.D., 1994, Nature 371, 757
\vskip 1truemm

Giallongo, E., Trevese, D., Vagnetti, F., 1991, ApJ 377, 345
\vskip 1truemm

Glazebrook, K., Ellis, R., Colless, M., Broadhurst, T.,

\hskip 2truemm Allington-Smith, J., Tanvir, Nial, 

\hskip 2truemm 1995, MNRAS 273, 157
\vskip 1truemm

Hawkins, M.R.S., Stewart, N.J., 1981, ApJ 251, 1
\vskip 1truemm

Helbig, P., Kayser, R., 1995, SISSA preprint
\vskip 1truemm

Hewett, P.C., Foltz, C.B., Chaffee, F.H., 1995, \par
\hskip 2truemm Astron. J. 109, 4, 1498
\vskip 1truemm

Hook, I.M., McMahon, R.G., Patnaik, A.R., Browne, I.W.A., \par
\hskip 2truemm Wilkinson, P.N., Irwin, M.J., Hazard, C., 1995 \par
\hskip 2truemm MNRAS 273, L63
\vskip 1truemm

Hudson, M.J., Dekel, A., Courteau, S., Faber, S.M., Willick, J.A., \par
\hskip 2truemm MNRAS 274, 305
\vskip 1truemm

Kassiola, A., Mathez, G., 1991, AA 230, 255
\vskip 1truemm

Kochanek, C.S., 1992, ApJ 384, 385
\vskip 1truemm

Jaffe, A., 1995, SISSA preprint

Jing, Y.P., Mo H.J., B\" \o rner, G., Fang, L.Z. 

\hskip 2truemm MNRAS: SISSA preprint
\vskip 1truemm

Lampton, M., Margon, B., Bowyer, S., 1976, ApJ 208, 177
\vskip 1truemm

Leonard, S., Lake, K., 1995, SISSA preprint
\vskip 1truemm

Llorente, A., P\'erez-Mercader, J., 1995, SISSA preprint
\vskip 1truemm

Loh, E.D., Spillar, E.J., 1986, ApJ 307, L1
\vskip 1truemm

Malhotra, S., Turner, E.L., 1994, SISSA preprint
\vskip 1truemm

Marshall, H.L., Avni, Y., Braccesi, A., Huchra, J.P.,

\hskip 2truemm Tananbaum, H., Zamorani, G., Zitelli, V.

\hskip 2truemm 1984, ApJ 283, 50
\vskip 1truemm

Mathez, G., 1976, AA 53, 15
\vskip 1truemm

Mathez, G., 1978, AA 68, 17
\vskip 1truemm

Mathez, G., Kassiola, A., Lachi\`eze-Rey, M., 1991, AA 242, 13
\vskip 1truemm

Mathez, G., Mellier, Y., Picat, J.P., 1995a, proceedings of \par
\hskip 2truemm ``Science with the VLT", ESO Workshop,  \par
\hskip 2truemm J. Walsh ed., M\" unich 1994.
\vskip 1truemm

Mathez, G., Van Waerbeke, L., Bonnet, H., 1995b \par
\hskip 2truemm submitted (Paper II), 
\vskip 1truemm

Nottale, L., 1993, Fractal Space-Time and Microphysics,

\hskip 2truemm chpt 7, World Scientific ed.
\vskip 1truemm

Phillipps, S., 1994, MN 269, 1077
\vskip 1truemm

Pierce, M.J., Welch, D.L., McLure, R.D., van den Bergh, 

\hskip 2truemm S., Racine, R., Stetson, P.B., 1994, Nature 371, 385
\vskip 1truemm

Press, W.H., Teukolsky, S.A., Vetterling, W.T., Flannery, B.P.,

\hskip 2truemm 1992, Numerical Recipes, Cambridge University Press,
p.686
\vskip 1truemm

Primack J.R., 1995, SISSA preprint 
\vskip 1truemm

Robin, A., 1989, PhD. dissertation, Universit\'e de Besan\c con
\vskip 1truemm

Schade, D., Hartwick F.D.A., 1994, ApJ 423, L85
\vskip 1truemm

Schmidt, M., 1968, ApJ 151, 393
\vskip 1truemm

Schmidt, M., 1972, ApJ 176, 273
\vskip 1truemm

Schmidt, M., Green, R.F., 1983, ApJ 269, 352
\vskip 1truemm

Shanks, T., Boyle, B.J., 1994, MN 271, 753
\vskip 1truemm

Shaver, P.A., ``High Redshift Quasars", to appear in \par
\hskip 2truemm {\it 17th Texas Symposium on Relativistic Astrophysics \par
\hskip 2truemm and Cosmology}, B\"ohringer et al. eds, Ann. New York \par
\hskip 2truemm Academy of Science
\vskip 1truemm

Turner, E.L., 1979, ApJ. 230, 291
\vskip 1truemm

Van den Bergh, S., 1994, PASP 106, 1113
\vskip 1truemm

V\'eron, P., 1983, Proc. 24th Li\`ege Astrophys. Col. \par
\hskip 3truemm `Quasars and Gravitational Lenses', Institut \par 
\hskip 3truemm d'Astrophysique, Univ. de Li\`ege, p.210 \par
\vskip 1truemm

V\'eron--Cetty, M.P., V\'eron, P., 1991, ESO Scientific
Report, p.8
\vskip 1truemm

Webster, R., 1994, XXIInd IAU General Assembly, The Hague.
\vskip 1truemm

Weinberg, S., 1989, Rev. Mod. Phys. 61, 1
\vskip 1truemm

Yoshii, Y., Peterson, B.A., 1995, ApJ 444, 15
\vskip 1truemm

Zaritski, D., White, S.D.M., 1993, ApJ 405, 464
\vskip 1truemm

Zieba S., Chyzy K., 1991, AA 241, 22

\end{document}